\documentclass[a4paper,12pt]{article}
\begin{document}
\title{Pitfalls of the theory of Bose-Einstein correlations in multiple particle production processes.}
\author{Kacper Zalewski\\
M. Smoluchowski Institute of Physics, Jagellonian University\\ and\\ Institute
of Nuclear Physics, Krak\'ow, Poland}
\maketitle

\begin{abstract}
Some basic difficulties and ambiguities on the way from the measured momentum
distributions to the inferred properties of the interaction regions are
reviewed and discussed.
\end{abstract}

\section{Introduction}

The Bose-Einstein correlations in multiple particle production processes are
studied mainly in order to get information about the interaction regions, i.e.
about the regions where the final state hadrons are created, and about the
hadron's last scattering surfaces which by definition are the boundaries in
space-time of the interaction regions. There is a number of things one would
like to know about an interaction region: its size and shape, its orientation
with respect to the event axis and in the case of the heavy ion collisions also
with respect to the impact parameter vector; the flows of matter, if any,
within the region; the nature of the matter filling the interaction region, its
equation of state and its phase transitions, if any. This list could be made
longer. One could try to obtain this information either from first principles,
or using phenomenological models, or in some model independent way from the
data. The approach from first principles is for the moment too difficult. In
the present paper we show some of the difficulties in deriving information
about the interaction region directly from the data without model assumptions.
The conclusion is that little reliable information can be derived without a
trustworthy model. The data in practice means momentum distributions: single
particle, two-particle, three-particle etc. There is some additional
information, e.g. every experiment gives some information about the positions
of the interaction vertices, but when we are interested in distances of the
order of tens of fermis or less this additional information does not contribute
significantly. The $k$-particle momentum distribution is given by the diagonal
elements of the $k$-particle density matrix in the momentum representation.
Most models assume that this can be expressed as a symmetrized product of
single particle density matrices \cite{KAR}:

\begin{equation}\label{karczm}
  \rho(\textbf{p}_1,\ldots,\textbf{p}_k;\textbf{p}_1,\ldots,\textbf{p}_k) = \sum_P \prod_{j=1}^k
  \rho_1(\textbf{p}_j;\textbf{p}_{Pj}),
\end{equation}
where the summation is over all the permutations of the momenta
$\textbf{p}_1,\ldots,\textbf{p}_k$. Thus, e.g. the single particle momentum
distribution is $\rho_1(\textbf{p};\textbf{p})$ and for the two-particle
distribution we get the well-known formula

\begin{equation}\label{}
  \rho(\textbf{p}_1,\textbf{p}_2;\textbf{p}_1,\textbf{p}_2) =
  \rho_1(\textbf{p}_1;\textbf{p}_1)\rho_1(\textbf{p}_2;\textbf{p}_2) + |\rho_1(\textbf{p}_1;\textbf{p}_2)|^2.
\end{equation}

\section{Emission function}

A popular strategy when studying the Bose-Einstein correlations is to evaluate
the emission function $S(K,X)$ (cf. \cite{WIH} and references given there)
related to the single particle density matrix by the formula

\begin{equation}\label{emifun}
   \rho_1(\textbf{p}_1;\textbf{p}_2) = \int\!\!\;dXS(K,X) e^{iqX}.
\end{equation}
Here $K$, $q$ and $X$ are four-vectors defined by

\begin{equation}\label{}
  K = \frac{1}{2}(p_1 + p_2);\qquad q = p_1 - p_2;\qquad X = \frac{1}{2}(x_1 +
  x_2).
\end{equation}
Note that $Kq = 0$. It is usual to interpret $S(X,K)$, for given $K_0$ and $x_0
= t$, in the spirit of the interpretation of the Wigner function, as the phase
space distribution of particles. Then $S(K,X)$ gives directly all the
information about the geometry and the evolution in time of the interaction
region. Moreover, it gives the momentum distribution of the produced particles
and information about the momentum-position correlations. The emission function
can be calculated when the production amplitudes are known \cite{SHU},
\cite{PCZ}. Let us consider now the prospects for calculating the emission
function from the data using formula (\ref{emifun}).

On both sides of equation (\ref{emifun}) the four-vector $K$ has the same,
fixed value. The right-hand-side is a Fourier transform of $S$ changing the
variables $X^\mu$ into the variables $q^\mu$. This Fourier transformation,
however, cannot be inverted, because the left hand side is known only for the
four-vectors $q$ satisfying the condition $Kq=0$, i.e. for only one value of
$q_0$ at each $\textbf{q}$. This implies that for any given density matrix
$\rho_1$ there is an infinity of different emission functions satisfying
relation (\ref{emifun}). They correspond to the infinite variety of possible
continuations of function $\rho(p_1;p_2)$ to unphysical values of $q_0$. The
data without additional information from theory, or from a model, are unable to
tell us which of these emission functions is the good one. In order to see how
this affects the conclusions about the interaction region let us consider the
following example.

We replace equation (\ref{emifun}) by the equation

\begin{equation}\label{}
 \rho_1(\textbf{p}_1;\textbf{p}_2) = \int\!\!\;dX'S(K,X') e^{iq(X' - c K)},
\end{equation}
where $c$ is an arbitrary real constant. This is, of course, legitimate,
because besides renaming the integration variable $X'$ we have just added to
the exponent $icKq = 0$. Changing the integration variable to $X = X' -cK$ we
get

\begin{equation}\label{}
\rho_1(\textbf{p}_1;\textbf{p}_2) = \int\!\!\;dXS(K,X + cK) e^{iqX}.
\end{equation}
Thus $S(K,X+cK)$ is another solution of equation (\ref{emifun}) for the same
density matrix $\rho_1$. The space-time distribution of the sources (more
precisely of the particles produced by the sources) is obtained by integrating
the emission function over $K$. Suppose now that for $c=0$ there are no $X$ ---
$K$ correlations, i.e. for every $K$ the space time region occupied by the
interaction region is the same. Then the $K$ integration does not affect the
space time distribution --- one can say that the interaction regions for all
the $K$-s are piled on top of each other. When $c$ becomes different from zero
the interaction regions corresponding to the different $K$ values get different
shifts. As a result the size of the overall space time region occupied by the
interaction region increases. With sufficiently large $c$ it can be made as
large as one wishes. For instance, one could make it the size of a football, or
living for an hour. Such extreme choices of $c$ can be eliminated using what we
know about the vertex positions in space-time, but using momentum distributions
only, they are just as good as $c=0$!

The obvious question is: what information besides the momentum measurements is
necessary to obtain reasonably accurate descriptions of the interaction
regions? One could try to evade the problem by saying: just forget about the
emission functions and your problem will disappear. We will show in the next
section that this is not a satisfactory answer.

\section{Density matrix and Wigner function}

Let us assume that all the hadrons have been produced in their final states
(scattered for the last time) instantaneously and simultaneously at some time
which we may chose as $t = 0$. Then, for any $t > 0$, all these hadrons can be
described by standard, time independent (in the interaction representation)
density matrices. According to assumption (\ref{karczm}) all these density
matrices can be expressed in terms of the single particle density matrix
$\rho_1$. Our key observation \cite{BIZ} is that replacing $\rho_1$ by

\begin{equation}\label{}
   \rho_{1\alpha}(\textbf{p}_1;\textbf{p}_2) =
  e^{i\alpha(\textbf{p}_1)}\rho_1(\textbf{p}_1;\textbf{p}_2)e^{-i\alpha(\textbf{p}_2)},
\end{equation}
where $\alpha(\textbf{p})$ is any real-valued function of momentum, does not
change the momentum distributions. This is obvious for $k=1$ and $k=2$. In
general, however, on the right-hand side of equation (\ref{karczm}), in every
monomial every momentum $\textbf{p}_i$ occurs exactly once as the first
argument of $\rho_1$ and exactly once as the second argument of $\rho_1$. Thus,
when $\rho_1$ gets replaced by $\rho_{1\alpha}$, it brings two factors
$e^{i\alpha(\textbf{p}_1)}$ and $e^{-i\alpha(\textbf{p}_1)}$ which cancel.

As a corollary let us note that this makes the definition of the emission
function, if obtained from the data using equation (\ref{emifun}), even more
ambiguous. Not only  for given $\rho_1$ there is an infinity of different
solutions for $S$, but moreover $\rho_1$ can be replaced by any of the infinity
of functions $\rho_{1\alpha}$ without affecting the fits to the experimental
data. Using the density matrix formalism instead of the emission functions, we
have the second ambiguity, but not the first. Let us see how it affects the
inferences concerning the interaction regions.

Let us make the simplifying assumption that

\begin{equation}\label{}
  \rho_1(\textbf{p}_1;\textbf{p}_2) =
  \frac{1}{\left(\sqrt{2\pi\Delta^2}\right)^3}\exp\left[-\frac{\textbf{K}^2}{2\Delta^2} -
  \frac{1}{2} R^2 \textbf{q}^2\right].
\end{equation}
Using the general formulae for the diagonal elements of the density matrix in
the coordinate representation and for the Wigner function in terms of the
elements of the density matrix in the momentum representation:

\begin{eqnarray}\label{}
\tilde{\rho}_1(\textbf{x};\textbf{x}) &=& \int\!\!dKdq\; e^{i\textbf{q}\cdot \textbf{X}}\rho_1(\textbf{p}_1;\textbf{p}_2),\\
W(\textbf{K},\textbf{X}) &=& \int\!\!\frac{dq}{(2\pi)^3} e^{i\textbf{q}\cdot
\textbf{X}}\rho_1(\textbf{p}_1;\textbf{p}_2)
\end{eqnarray}
one gets

\begin{eqnarray}\label{}
\tilde{\rho}_1(\textbf{x};\textbf{x}) &=&
\frac{1}{\sqrt{2\pi R^2}^3}\exp\left[-\frac{\textbf{x}^2}{2R^2}\right], \\
W(\textbf{K},\textbf{X}) &=& \frac{1}{(2\pi R\Delta)^3}\exp
\left[-\frac{\textbf{K}^2}{2\Delta^2} - \frac{\textbf{X}^2}{2 R^2}\right].
\end{eqnarray}
The diagonal elements of the density matrix in the coordinate representation
yield the distribution of particles in space at $t=0$ (keep in mind that we are
using the interaction representation). This is a spherically symmetric Gaussian
with the root mean square width in every coordinate equal $R$. The Wigner
function is usually interpreted as an approximation to the phase space
distribution. In our case it confirms the result for the space distribution
and,  moreover, gives the information that the momentum distribution is
uncorrelated to the space distribution and is also Gaussian with the root mean
square width $\Delta$.

Let us now replace the Gaussian $\rho_1$ by the corresponding $\rho_{1\alpha}$
with

\begin{equation}\label{}
  \alpha(\textbf{p}) = \frac{1}{2}c \textbf{p}^2,
\end{equation}
where $c$ is an arbitrary real number. A simple calculation gives

\begin{eqnarray}\label{}
\tilde{\rho}_{1\alpha}(\textbf{x};\textbf{x}) = \nonumber \\
\frac{1}{\sqrt{2\pi (R^2 + c^2\Delta^2)}^3}\exp\left[-\frac{\textbf{x}^2}{2(R^2 + c^2\Delta^2)}\right], \\
W_\alpha(\textbf{K},\textbf{X}) = \nonumber \\
\frac{1}{(2\pi R\Delta)^3}\exp \left[-\frac{\textbf{K}^2}{2\Delta^2} -
\frac{(\textbf{X + c K})^2}{2 R^2}\right].
\end{eqnarray}
The first formula shows that the distribution in space remains spherically
symmetric  and Gaussian, but the mean square radius can be any number not less
than $R$. The second equation shows that momentum-position correlations have
been introduced and that the smearing of the interaction region is due to a
mechanism similar to that described in the preceding section: the interaction
regions for particles with different values of the momentum $\textbf{K}$ shift
with respect to each other.

Note that, if we have a model which predicts unambiguously the density matrix
$\rho_1$, there is no problem. The problem arises if one insists that the
conclusions about the interaction region should be deduced from the data
without additional assumptions. An interesting question, however, is: suppose
that we have two models based on completely different physical assumptions, one
giving a matrix $\rho_1$ which agrees with experiment, and the other a
corresponding matrix $\rho_{1\alpha}$ which, as follows from the preceding
discussion, agrees with experiment just as well; how can we tell which of the
two models is the good one? Experimental data on momentum distributions cannot
help however precise they are. The standard advice to look at more particle
distributions is also useless.

\section{Distribution of distance between points where identical particles are produced}

Let us consider a pair of identical particles with total momentum $2\textbf{K}$
and denote the distance between their production points by $\textbf{x}$. An
important strategy is to study the distribution of the distances
($\textbf{x}$), i.e. ${\cal{S}}_\textbf{K}(\textbf{x})$ \cite{PCZ}, \cite{WIH},
\cite{LIS}. The relation of the function ${\cal S}_\textbf{K}$ to the data is

\begin{equation}\label{disdis}
 \frac{dN/dp_1dp_2}{dN/dp_1dN/dp_2} = \int\!\!d^3x{\cal S}_\textbf{K}(\textbf{x'})\left[|\phi_\textbf{q'}(\textbf{x'})|^2 -
 1\right],
\end{equation}
where the primed quantities are defined in the center of mass frame of the pair
and $\phi_\textbf{q'}(\textbf{x'})$ is the two-particle wave function of the
pair in its center of mass frame. When two particles are far from each other
their relative momentum is well defined and equals $\textbf{q'}$. In particular
for free particles the content of the square bracket equals $\cos(
\textbf{q'}\cdot \textbf{x'})$, but one can include final state interactions by
using more complicated wave functions.

The solution of equation (\ref{disdis}) for the free particle case can be
easily done by inverting the Fourier transformation. For more complicated cases
one has to use more advanced methods known as imaging \cite{BRD1}, \cite{BRD2},
\cite{BRD3}, but there is no difficulty of principle. In this approach function
${\cal S}_\textbf{K}$ is unambiguously given by the data. This is a nice change
after the two methods described in the preceding two sections and, therefore,
the method is rapidly gaining popularity.

Let us note, however, that the left hand side of equality (\ref{disdis}), in
approximation (\ref{karczm}) is

\begin{equation}\label{}
\frac{dN/dp_1dp_2}{dN/dp_1dN/dp_2} =
\frac{|\rho_1(p_1;p_2)|^2}{\rho_1(p_1;p_1)\rho_1(p_2;p_2)} + 1.
\end{equation}
The right hand side is clearly invariant with respect to the transformation
discussed in Section~3. Therefore, there is an ambiguity concerning the
conclusions about the interaction region as discussed there. We infer that
function ${\cal S}_\textbf{K}$ can be evaluated unambiguously from the data,
but its implications for the interaction region are highly ambiguous. Thus,
again there is no simple way from the data to the information about the
interaction region. It is plausible that each of the three approaches discussed
here, and certainly many others, share essentially the same difficulty, which
just changes its form when the formalism is being changed.

\section{Conclusions}

The analysis presented in the previous three sections strongly suggests that
little information about the interaction region can be obtained from the data
on the momentum distributions without making additional assumptions. Usually
this problem is solved by using a model. Every model contains assumptions
impossible to prove by just using the data on the momentum distributions.
Within the model these assumptions could be justified by arguments of beauty,
simplicity, or by invoking some general principles. This rises the following
important problem, however.

Suppose that two models differ only in these additional assumption and thus
they give exactly the same predictions for all the momentum distributions.
Suppose further that they correspond to widely different and conflicting
pictures of the interaction region and of what happens there. How should one
decide which model is the realistic one? This is not an academic problem. It
has been already noticed that widely different model can fit the same data, cf.
e.g. \cite{PAG}, \cite{BIA}.

\end{document}